\documentclass[preprint]{revtex4-1}
\usepackage{graphicx}
\usepackage{color}

\renewcommand\figurename{Figure}

\begin{document}

\title{Fractality \textit{\`a la carte}: a general particle-cluster aggregation model}
\author{J. R. Nicol\'as-Carlock}
\affiliation{Instituto de F\'isica, Benem\'erita Universidad Aut\'onoma de Puebla, Apdo. Postal. J-48, Puebla 72570, Mexico.}
\author{J. L. Carrillo-Estrada}
\email[Author to whom correspondence should be addressed: ]{carrillo@ifuap.buap.mx}
\affiliation{Instituto de F\'isica, Benem\'erita Universidad Aut\'onoma de Puebla, Apdo. Postal. J-48, Puebla 72570, Mexico.}
\author{V. Dossetti}
\affiliation{Instituto de F\'isica, Benem\'erita Universidad Aut\'onoma de Puebla, Apdo. Postal. J-48, Puebla 72570, Mexico.}


\begin{abstract}

{\bf
Aggregation phenomena are ubiquitous in nature, encompassing out-of-equilibrium processes of fractal pattern formation, important in many areas of science and technology \cite{vicsek1992book, meakin1998book, sander2000, sander2011}.
Despite their simplicity, foundational models such as diffusion-limited aggregation \cite{witten81} (DLA) or ballistic aggregation \cite{vold63} (BA), have contributed to reveal the most basic mechanisms that give origin to fractal structures. Hitherto, it has been commonly accepted that, in the absence of long-range particle-cluster interactions, the trajectories of aggregating particles, carrying the entropic information of the growing medium, are the main elements of the aggregation dynamics that determine the fractality and morphology of the aggregates \cite{meakin1984a, matsushita1986, huang1987, huang2001, ferreira2005}. 
However, when interactions are not negligible, fractality is enhanced or emerges from the screening effects generated by the aggregated particles, a fact that has led to believe that the main contribution to fractality and morphology is of an energetic character only, turning the entropic one of no special significance, to be considered just as an intrinsic stochastic element \cite{meakin1983b, nakagawa1992, block1991, vandewalle1995, pastor1995, indiveri1999, jullien1986a, meakin1988, meakin1991}.
Here we show that, even when long-range attractive interactions are considered, not only screening effects but also, in a very significant manner, particle trajectories themselves are the two fundamental ingredients that give rise to the fractality in aggregates. 
We found that, while the local morphology of the aggregates is determined by the interactions, their global aspect will exclusively depend on the particle trajectories. Thus, by considering an effective aggregation range, $\lambda$, we obtain a wide and versatile generalization of the DLA and BA models.
Furthermore, for the first time, we show how to generate a vast richness of natural-looking branching clusters with any prescribed fractal dimension $D\in[1,D_0]$, very precisely controlled, being $D_0$ the fractal dimension of clusters resulting from the underlying non-interactive aggregation-model, DLA or BA, used. 
}

\end{abstract}

\maketitle

Since the introduction of the DLA model, a plethora of studies and models have been developed, trying to understand the most basic mechanisms that give rise to self-affine structures in aggregation phenomena \cite{vicsek1992book, meakin1998book, sander2000, sander2011}. In particle-cluster aggregation, the paradigmatic DLA and BA models have shown that, if long-range interactions are negligible, the characteristics of the growing medium, such as its temperature and viscosity, encoded in the aggregating particle trajectories, will define the morphology and fractality of the growing cluster \cite{meakin1984a, matsushita1986}. This fact is remarkably exhibited when, by changing the particles mean squared displacement (MSD), a transition from compact BA clusters to branching DLA ones can be achieved \cite{huang1987, huang2001, ferreira2005}. On the other hand, when interactions can no longer be disregarded, aggregation dynamics could become quite complex. Nonetheless, experimental and computational models have shown that the effects of interactions on the morphology and fractality of clusters are mainly two. Short-range repulsive interactions decrease the fractality, as they allow particles to reach a minimum in the energetic landscape, thus producing compact clusters \cite{meakin1983b, nakagawa1992}. Conversely, long-range attractive interactions increase fractality, generating highly ramified clusters, due to screening effects coming from the interaction-range of the aggregated particles \cite{meakin1983b, block1991, vandewalle1995, pastor1995, indiveri1999, wen1997, kun1998, hastings2001, carlier2012}. However, one must carefully consider that the main mechanism responsible for the cluster's morphological and fractal features cannot trivially be attributed to a single process of an energetic or entropic nature, when the underlying aggregation dynamics comes up as a result of the interplay between the stochastic trajectories of the particles and the quasi-deterministic particle-cluster interactions \cite{pastor1995, block1991}. While several technically and conceptually simple approaches have regarded the energetic character of the dynamics as the most important factor to the clusters morphology and fractality \cite{meakin1983b, block1991, jullien1986a, meakin1988, meakin1991, nakagawa1992, indiveri1999}, in this Letter we show that the entropic contribution cannot be trivially considered as an intrinsic stochastic element of the system, but as an important aspect that not only contributes to the fractality of the clusters, but also as a remarkable source of diversity in fractal pattern formation.

\begin{figure*}[tb]
\includegraphics[width=\textwidth]{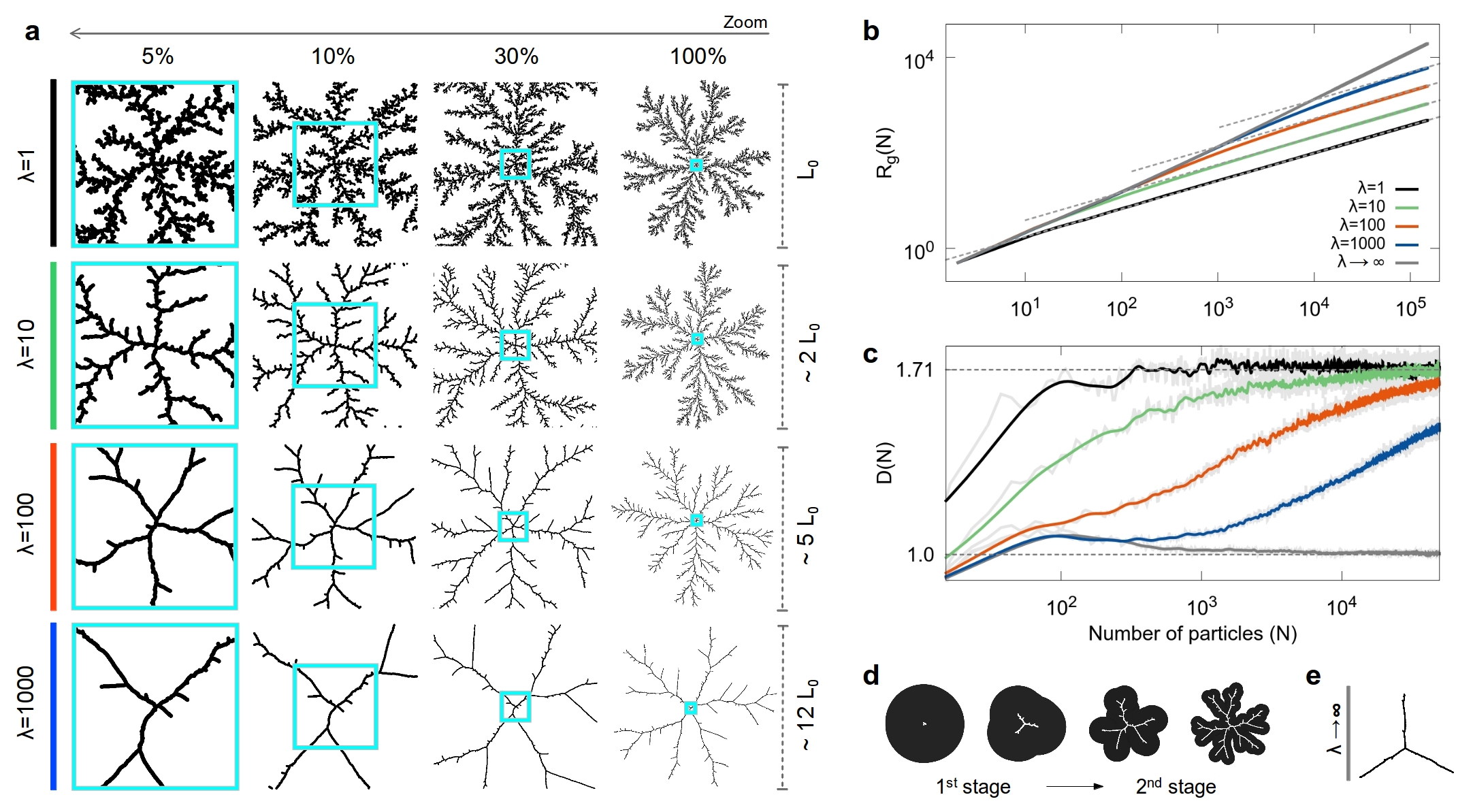}
\caption{\label{edla} \textbf{DLA-based multiscaling aggregates.} \textbf{a} Aggregates containing $N=150\times 10^3$ particles each, for $\lambda=1,10,100$ and $1000$ units, visualized at $5\%,10\%,30\%$ and $100\%$ of their total size. The blue squares display the multiscaling evolution of the structure. \textbf{b} Radius of gyration, $R_g$, and \textbf{c} fractal dimension, $D$, versus the number of aggregated-particles, $N$, in log-log and lin-log plots, respectively. Notice that, when $\lambda\to\infty$, the structure of the aggregates tends to MF $(D=1)$. Each curve was computed as an average over an ensemble of aggregates. \textbf{d} Evolution of the growing front for the first two stages of growth (see text). \textbf{e} Typical structure of a MF aggregate.}
\end{figure*}

To this end, we introduce a two-dimensional particle-cluster model that, trough the incorporation of an effective interaction or aggregation range $\lambda$, deeply generalizes and enhances the capabilities of the standard DLA and BA models. By exploring the aggregation dynamics under constant $\lambda$ and $\lambda=\lambda (N)$, where $N$ is the number of aggregated particles in the cluster, this simple but non-trivial stochastic scheme for aggregation allows us to separate and characterize the subtle contributions of energetic and entropic character of the dynamics to the fractality and morphology of the clusters, as well as it reveals features not previously seen before \cite{meakin1983b, grzegorczyk2004a}. The aggregation dynamics of this model is quite simple, assuming that (i) each aggregated particle in the cluster is assigned with an interaction radius $\lambda$, and that (ii) as soon as the path of a wandering particle intersects the cluster's interaction boundary, it moves radially and ballistically to the closest particle in the cluster, getting irreversibly and rigidly attached. Notice that the interaction boundary is defined by the overlap of the individual interaction regions of the aggregated particles (see Fig.\ 1\textbf{d} and 2\textbf{d}). Additionally, distance quantities are measured in particle-diameter units, which are fixed to one. In order to characterize the fractal properties of our aggregates, we used the radius of gyration, $R_g=k N^{\beta}$, where $k$ is a constant, $\beta=1/D$, $D$ being fractal dimension. Typically, $D$ is estimated from the slope of plots of $R_g$ vs $N$ in logarithmic scale, or equivalently (as in this work), through the numerical derivative of $\log R_g$ with respect to $N$ (see Methods section).

\begin{figure*}[tb]
\includegraphics[width=\textwidth]{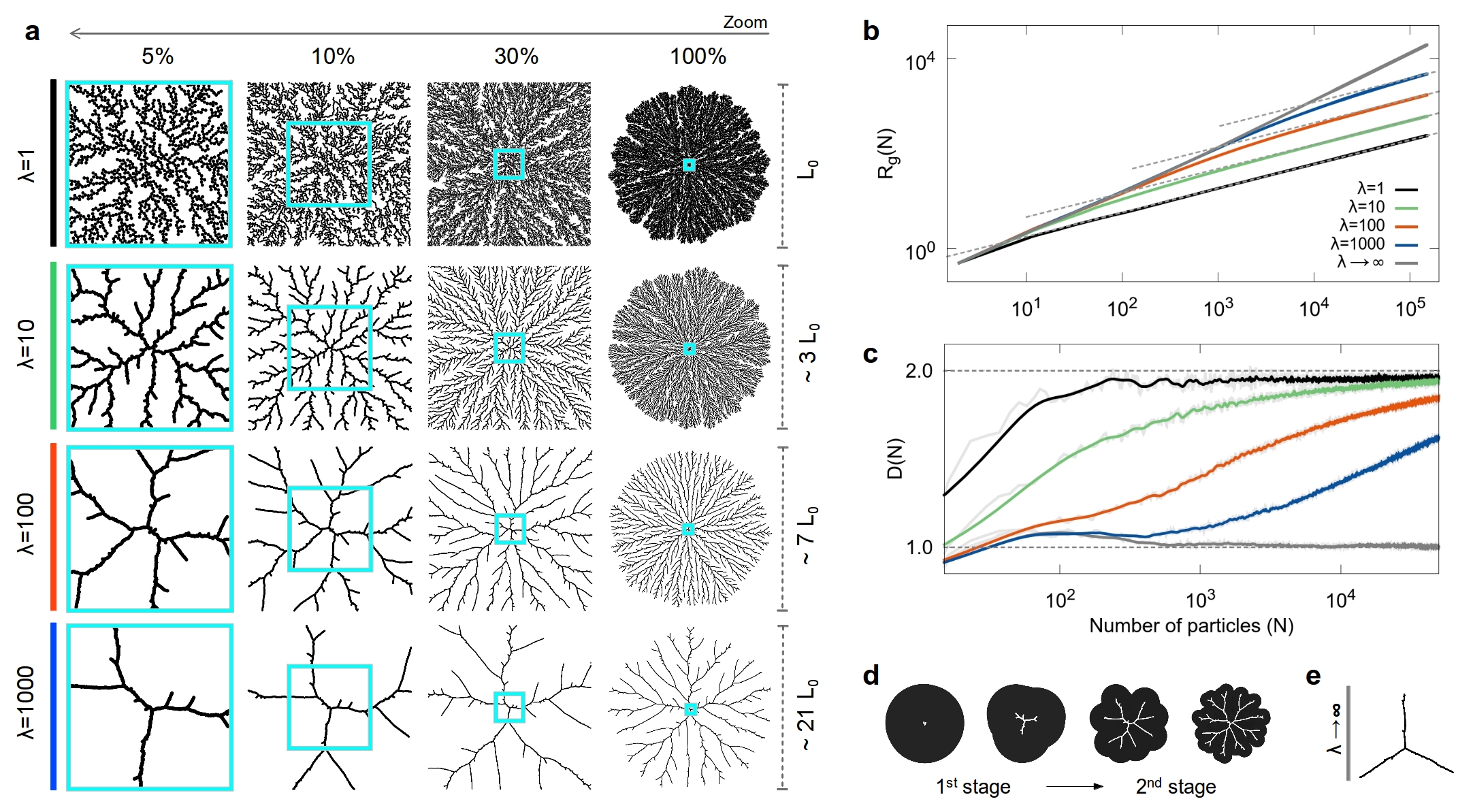}
\caption{\label{eba} \textbf{BA-based multiscaling aggregates.} \textbf{a} Aggregates containing $N=300\times 10^3$ particles each, for $\lambda=1,10,100$ and $1000$ units, visualized at $5\%,10\%,30\%$ and $100\%$ of their total size. The blue squares display the multiscaling evolution of the structure. \textbf{b} Radius of gyration, $R_g$, and \textbf{c} fractal dimension, $D$, versus the number of particles, $N$, in log-log and lin-log plots, respectively. Notice that, when $\lambda\to\infty$, the structure of the aggregates tends to MF $(D=1)$. Each curve was computed as an average over an ensemble of aggregates. \textbf{d} Evolution of the growing front for the first two stages of growth (see text). \textbf{e} Typical structure of a MF aggregate.}
\end{figure*}

Let us first discuss our results for the case of interactions with a constant range. For $\lambda=1$, or \emph{direct-contact} interaction, we have the usual DLA or BA models (see Figs.\ 1 and 2, respectively); we will refer to the known fractal dimension of the DLA or BA aggregates as $D_0=1.71$ or $2$, respectively. Further on, when $\lambda>1$, two main features will emerge due to the interplay of interactions and particle trajectories: a multiscaling branching growth and a crossover in fractality, from $D=1$ (as $\lambda \to \infty$) to $D=D_0$ (when $N \to \infty$). The former could be qualitatively characterized by three growth stages. In the first stage, when the radial size of the cluster is small compared to $\lambda$, the growth is limited by the interactions. The aggregated particles are close enough so that their individual interaction regions are highly overlapped, forming an almost circular envelope around the cluster, and making the last aggregated-particles the most probable aggregation points for particles incoming to that region. More likely, three arms grow, clearly seen as $\lambda \to \infty$. This stage is characterized by $D \approx 1$. In the second stage, the envelope starts to develop small deviations from its initially circular form, with typically three main elongations or growth instabilities, associated with the main branches. When the distance between the tips of two adjacent branches becomes of the order of $\lambda$, a bifurcation process starts, generating \emph{multiscaling} growth. After that, when the interactive envelope develops a branched structure itself (see Fig.\ 1\textbf{d} and 2\textbf{d}), particles are able to penetrate into the inner regions of the aggregate and a transition in growth dynamics takes place, from \emph{interaction-limited} to \emph{trajectory-limited}. In the third stage, when the distance among the tips of the main branches becomes much larger than $\lambda$, growth is limited by the mean squared displacement of the wandering particles. This is clearly appreciated for $\lambda=100$ and $1000$ in Figs.\ 1 and 2. For $\lambda=10$, even though the same crossover is present, it is masked by the \emph{small-cluster}-size regime. Notice that in both DLA and BA cases, for a larger interaction range, a higher number of aggregated-particles is required to recover the asymptotic behavior. Yet, the asymptotic value $D=D_0$ and the cluster's global structure characteristic of DLA or BA will be recovered as $N\to\infty$. However, as $\lambda\to\infty$, the structural and fractal features will tend towards a mean-field (MF) behavior \cite{jullien1986a}, characterized by $D=1$ and three well defined branches despite the particle trajectories.

\begin{figure*}[tb]
\includegraphics[width=\textwidth]{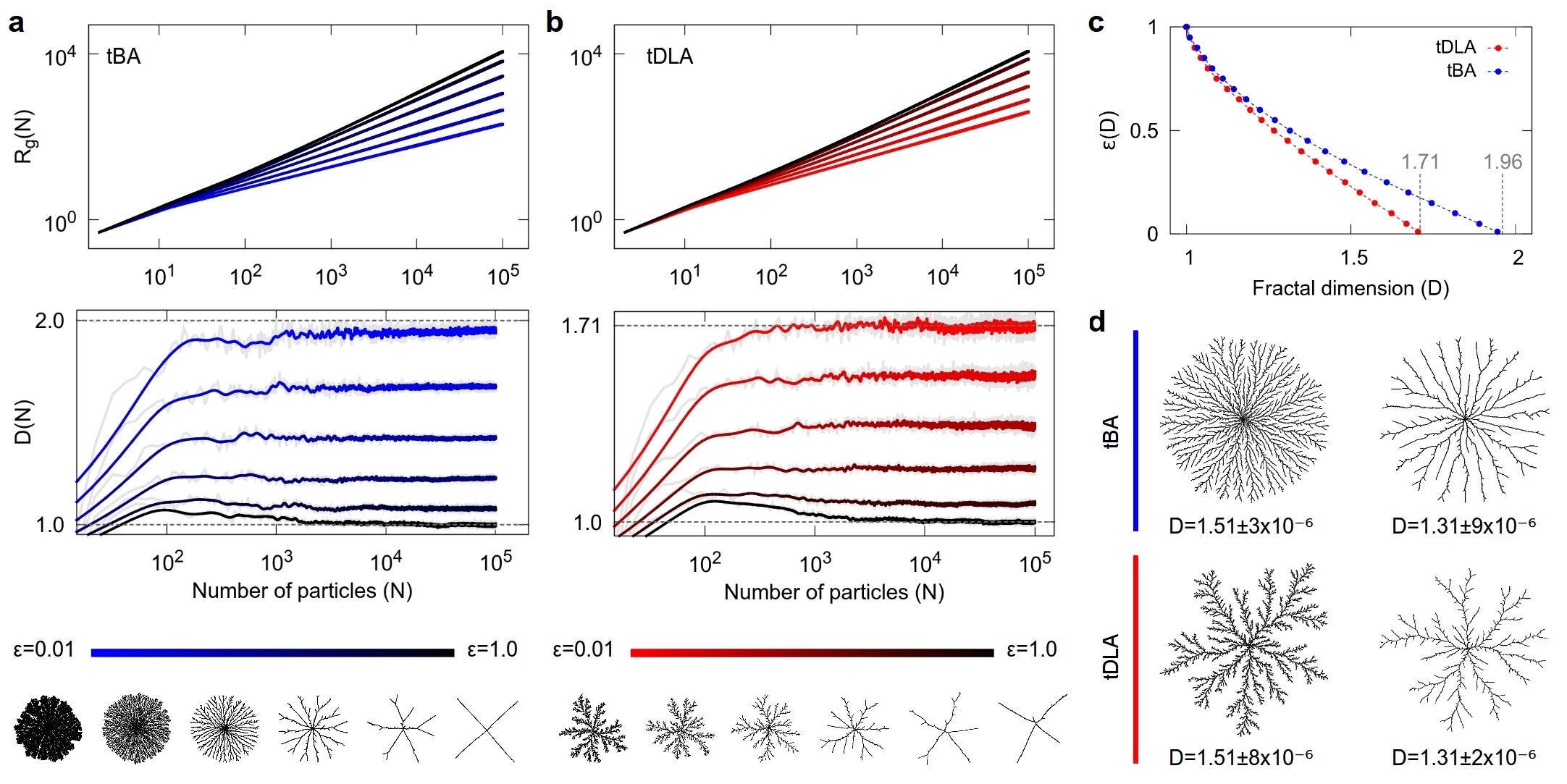}
\caption{\label{control} \textbf{Tunable aggregates.} In \textbf{a} and \textbf{b}, log-log plots for $R_g$ and lin-log for $D$ vs $N$, for aggregates grown with specific values of $\varepsilon$ in the interval $[0.01,1]$, for BA (blue) and DLA (red) up to $N=10^5$ particles. Notice how the multiscaling behavior gives way to a single well-defined fractal dimension $D=D(\varepsilon)$. At the bottom, one can appreciate the transition in these monofractal-aggregates morphology with respect to $\varepsilon$ . In \textbf{c}, plots of $\varepsilon$ vs $D$ for aggregates based on DLA and BA. These numerically obtained curves can be used to grow clusters with any prescribed $D$. \textbf{d} BA- and DLA-based clusters are shown with the same fractal dimension, $D=1.51$ and $1.31$, grown with a very high precision around de desired value.}
\end{figure*}

\begin{figure}[tb]
\includegraphics[width=0.6\textwidth]{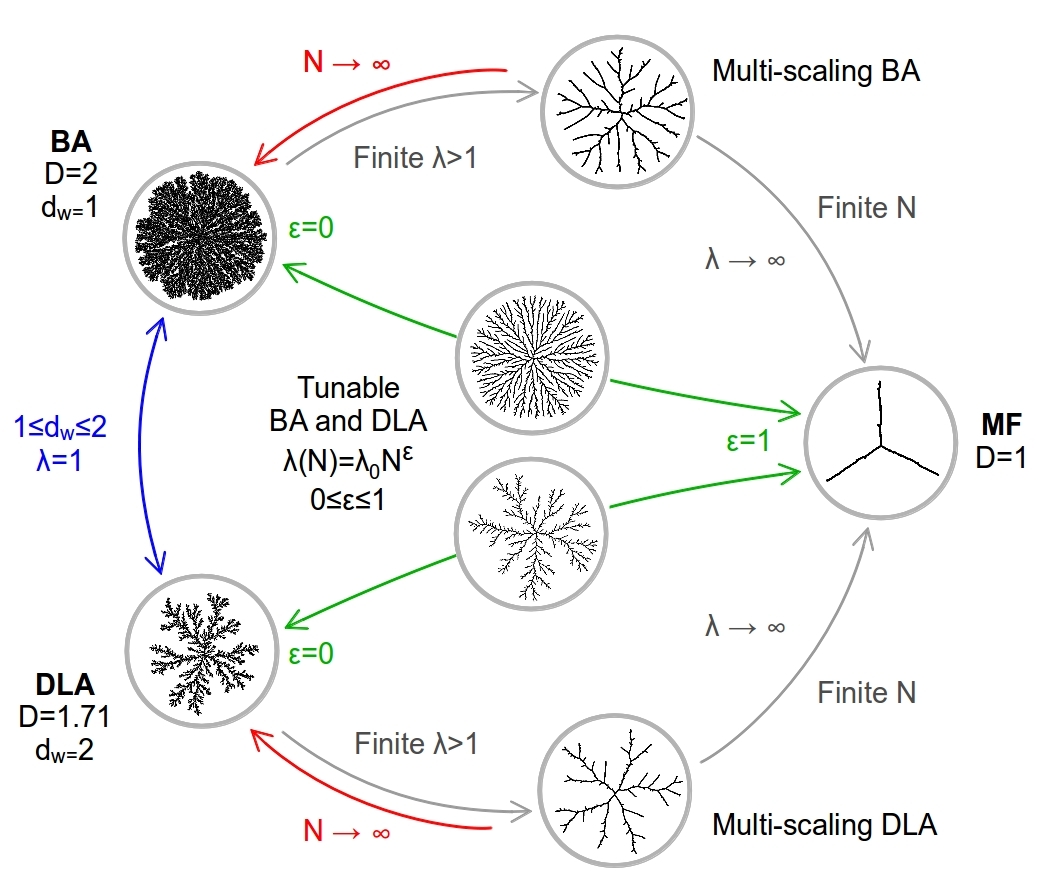}
\caption{\label{diagram} \textbf{Fractality diagram.} The diversity of aggregates that can be obtained with the introduction of $\lambda(N)$ is astounding. (\emph{i}) With $\varepsilon=0$ and $\lambda_0=1$, one has the well known transition from BA to DLA by changing the trajectories' fractal dimension, $d_w$, from to 1 to 2, respectively. (\emph{ii}) With $\varepsilon=0$ and a constant $\lambda_0>1$ multiscaling aggregates are obtained, while MF behavior is obtained in the limit $\lambda \rightarrow \infty$ for a finite $N$. Otherwise, with a large enough $N$, we get back to the usual BA or DLA behavior. Finally, (\emph{iii}) with $\varepsilon>0$ and $\lambda_0=1$, aggregates with a tuned $D$ from BA or DLA to MF, can be obtained by adequately scaling $\lambda$ with $N$.}
\end{figure}

The previous results have two important consequences. Firstly, even though interactions leave a strong print in the local cluster's structure and fractality, the stochastic nature of the particle trajectories will ultimately determine their global characteristics. Nonetheless, clusters cannot be trivially described by a single fractal dimension as it was previously thought \cite{meakin1983b, grzegorczyk2004a}, since their multiscaling behavior is able to span many orders of magnitude in the space occupied by the clusters. Secondly, when long-range attractive interactions are introduced in the growth dynamics, the only way to obtain monofractal clusters, with a well defined fractal dimension (smaller than $D_0$), i.e., with serf-affinity, is to maintain a proper balance of energetic and entropic contributions to the fractality. In fact, taking into account that the size of the clusters in space is proportional to the radius of gyration $R_g \propto N^{1/D}$, this balance can be achieved by scaling the interaction range with the number of particles in the cluster, through the generalized $\lambda(N)=\lambda_0N^\varepsilon$. Here, $\lambda_0$ is fixed to one, while $\epsilon$ is the scaling parameter that takes values in $[0,1]$. As shown in Fig.\ 3, this approach precisely and uniquely defines $D$ for each given value of $\varepsilon$. In addition, using the underlying DLA and BA models, within this scheme (scaling $\lambda$ with $N$), allows us to generate a rich variety of appealing and novel \emph{tunable} aggregates with $D$ in $[1,D_0]$, not previously seen \cite{meakin1985, filippov2000}. Yet, even in this case, the overall morphology of the cluster is still defined by the MSD of the wandering particles.

It is worth to mention that, in contrast with \emph{screened-growth} or \emph{sequential-algorithm} models \cite{meakin1985, filippov2000}, the fractality displayed by the clusters generated with our model is an emergent property of the aggregation dynamics. Additionally, our model remains simple in contrast to \emph{Lagrangian} models (i.e., based on a molecular dynamics approach), where the cluster fractality comes from the forces (stochastic and deterministic) experienced by the particles \cite{nakagawa1992, block1991}. Moreover, our results allow us to understand, in a very clear manner, the effects coming from the wandering-particles' MSD and from the interactions themselves, on the clusters' morphology and fractality.

As final remarks, Fig.\ 4 shows the whole family of fractal aggregates that can be generated under this simple scheme. When the interaction range is kept constant, all multiscaling aggregates will belong to the same universality class corresponding to the underlying non-interactive aggregation model used to generate them. On the other hand, when the interaction range is properly scaled with the cluster's size, $N$, the full range of tunable monofractal aggregates can be generated with $D \in [1,2]$. Also notice that our model can easily be extended to dimensions higher than 2. Therefore, these very important features of our model can be exploited beyond aggregation phenomena. We anticipate our findings will provide important insights into the study of fractal growth phenomena, from networks \cite{aono15} and branching morphogenesis \cite{affolter09}, to bio-inspired materials engineering \cite{ziaei15}, among others.

\section*{Methods}

{\bf Attractive interactions.} In order to introduce attractive interactions in our model, an interaction circle of radius $\lambda$ is assigned to each aggregated-particle, centered around the particle. This effective interaction distance, $\lambda$, is measured in particle-diameter units, acquiring values $\lambda \geq 1$. Additionally, as explained before, It can remain constant all along the dynamics or scale with $N$, depending on the desired application, i.e., to produce a multiscaling aggregate or an aggregate with tuned fractal dimensions. Further on, for aggregates based on BA, we follow the standard procedure in which particles are launched at random from the circumference of a circle of radius $L = r_{max} + \delta$, with equal probability in position and direction of motion \cite{ferreira2005}. Here, $r_{max}$ is the distance of the farthest particle in the cluster with respect to the seed particle placed at the origin. In our simulations we used $\delta=1000$ particle diameters to avoid undesirable screening effects. In the case of aggregates based on DLA, particles were launched from a circle of radius $L = r_{max} + \lambda + \delta$, with $\delta=100$. The mean free path for the motion of the particles is then set to one particle diameter, $\lambda_0 = 1$. We also used a scheme that modifies (increments) the mean free path as the particles wander at a distances greater than $L$ or in between branches, and set a killing radius at $L_k = 2L$, in order to speed up the aggregation process. In both scenarios, when a particle crosses for the first time the aggregation zone of any particle, characterized by $\lambda$, it moves radially and ballistically the closest particle along its path in an irreversible (no disaggregation) and rigid (no re-accommodation) form, as schematized step-by-step in Extended Data Figure 1.

{\bf Computing the fractal dimension.} The fractal dimension, $D$, is estimated from the radius of gyration, $R_g$, by means of a linear fit of the function $R_g=kN^{\beta}$ to the numerical data in a log-log plot, where $k$ is a constant, $N$ is the number of aggregated particles and $\beta=1/D$. In practice, it is assumed that $\beta$ is constant as long as the number of particles in the cluster is large. Because the multiscaling models do not develop a constant fractal dimension at their early stages of growth, the simplest way to quantitative and qualitative measure the behavior of $D$ is through the derivative of $R_g$ in the logarithmic scale. We did so by means of standard two and three point numerical differentiation methods: $f'(x)=[f(x+h)-f(x)]/h$, at the ends of the differentiation intervals and $f'(x)=[f(x+h)-f(x-h)]/2h$, in between. Here $f(x)=\log R_g(N)$ and $R(g)$ is computed as a discrete quantity therefore, $h$ is set as the distance between the points, $x=x_j$ and $x+h=x_{j+1}$. In all cases, $R_g$ is computed as an average over a large ensemble of aggregates. Specifically, the results for the multiscaling aggregates (i.e., with $\lambda$ kept constant) shown in Figures 1\textbf{a}, 1\textbf{b}, 2\textbf{a}, and 2\textbf{b}, were obtained over 64 and 15 aggregates containing $1.5\times10^5$ and $3\times10^5$ particles for those based on DLA and BA, respectively. In this case, $R_g$ was calculated every 10 particles. In Figures 1\textbf{c} and 2\textbf{c}, 192 and 128 aggregates containing $5\times10^4$ and $10^5$ particles were used to obtain de averages, respectively, while $R_g$ was calculated every 7 particles in order to capture the features of $D$ at small scales in $N$. The results for tunable aggregates based on DLA and BA (with $\lambda = \lambda(N)$), shown in Figure 3, were obtained over 128 and 48 aggregates, respectively, containing $10^5$ particles, and $R_g$ was computed every 10 particles. We must point out that the fluctuations observed in $D$, depicted by the grey curves in Figures 1\textbf{c}, 2\textbf{c}, 3\textbf{a} and 3\textbf{b}, are due to the numerical and local aspect of the derivative's estimation and the stochasticity of the model. These fluctuations decrease as $N$ increases, when the aggregates tend towards a more well defined structure. Thus, with the purpose to improve the visualization of the observed tendencies, the curves in color were computed by means of a running-average over $N$.

{\bf Tunable aggregates.} Aggregates with a prescribed (tuned) fractal dimension, either based on BA or DLA, have a well defined $D$ for each given value of $\varepsilon$. Therefore, we computed $D$ for $\varepsilon \in \{0.01,0.05,0.1,...,0.95,1.0\}$ in order to obtain the functional dependence of $\varepsilon$ on $D$. Then, for a desired $D^*$, we estimated the corresponding value of $\varepsilon$ through a linear approximation using the two closest points in $D$ to the desired $D^*$. In this case, a linear approximation is a valid method since the difference among consecutive points in $\varepsilon(D)$ is small and the curve is well behaved as can be appreciated in Figure 4\textbf{c}. 

\clearpage

\section*{Extended Data}

\begin{figure}[h]
\renewcommand\figurename{Extended Data Figure}
\setcounter{figure}{0}
\includegraphics[width=3.5in]{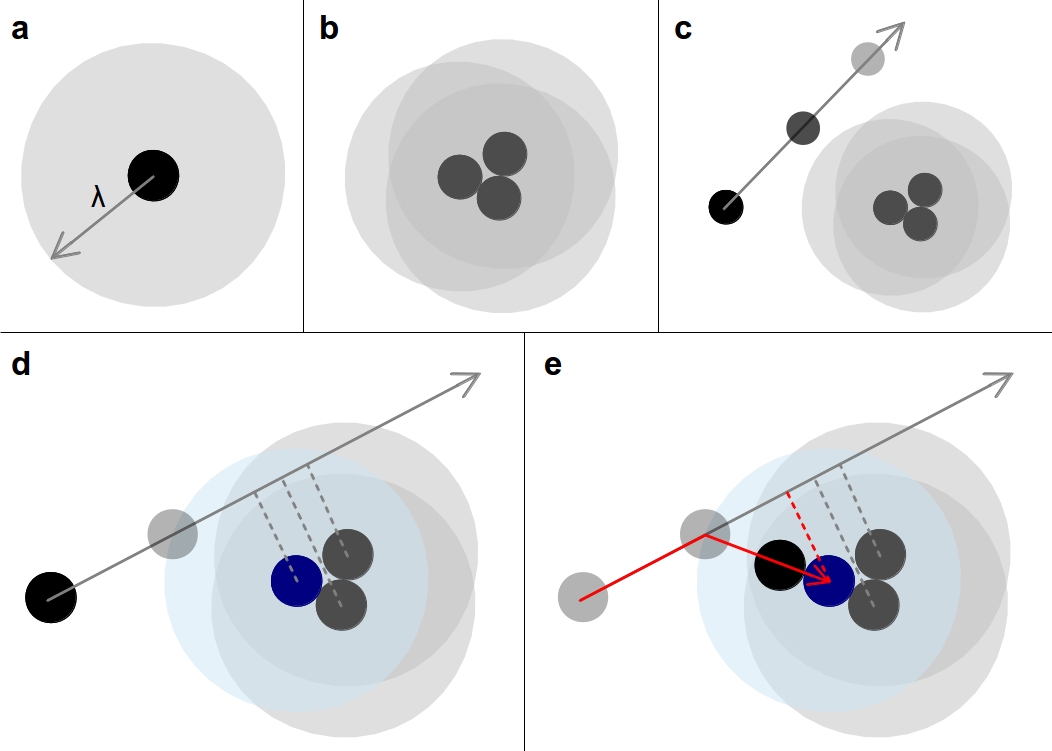}
\caption{\label{model} \textbf{Implementation of the interaction.} A step-by-step diagram is provided regarding the implementation of the attractive interactions in our model. \textbf{a} Each particle in the cluster is assigned an effective radius of aggregation $\lambda$, starting with the seed particle. \textbf{b} Upon aggregation, the independent interaction regions of each particle, defined by $\lambda$, overlap among them. \textbf{c} A particle far from this region does not interact with the cluster until \textbf{d} its trajectory is such that, for its next step, it intersects for the first time the interaction boundary of any aggregated-particle. This is determined when its perpendicular distance to the particles in the cluster is less than $\lambda$. Then, the position of the aggregated-particles are projected to determine the closest one along the direction of motion. Finally, \textbf{e} the position of first crossing is computed and the position of the new particle in the cluster is determined.}
\end{figure}

\vspace{7mm}

{\bf Supplementary Information} is available in the online version of the paper.\\

{\bf Acknowledgments} The authors gratefully acknowledge the computing time granted on the super\-com\-pu\-ters MIZTLI (DGTIC-UNAM) and, through the project ``Cosmolog\'{\i}a y as\-tro\-f\'{\i}\-si\-ca relativista: objetos compactos y materia obs\-cu\-ra'', on \hbox{XIUHCOATL} (\hbox{CINVESTAV}). We also acknowledge partial financial support from CONACyT and from VIEP-BUAP.\\

{\bf Author Contributions} J.R.N.C.\ carried out all the calculations, performed the analysis of the data and prepared all the figures. V.D.\ supervised the development of the calculations. J.L.C.E.\ supervised the research. All of the authors contributed in the discussion of the results and the preparation of the manuscript.\\

\end{document}